# Influence of microstructure and atomic-scale chemistry on iron ore reduction with hydrogen at 700°C

Se-Ho Kim[1,**], Xue Zhang[1,**], Kevin Schweinar[1], Isnaldi R. Souza Filho[1], Katja Angerendt[1], Yan Ma[1], Dirk Vogel[1], Leigh T. Stephenson[1], Ayman A. El-Zoka[1], Jaber Rezaei Mianroodi[1], Michael Rohwerder[1], Baptiste Gault[1,2,*], Dierk Raabe[1,*]

[1] Max-Planck-Institut für Eisenforschung, Max-Planck-Straße 1, 40237 Düsseldorf, Germany

[2] Department of Materials, Imperial College, South Kensington, London SW7 2AZ, UK

*corr. Authors: d.raabe@mpie.de, b.gault@mpie.de

** Kim and Zhang both act as first co-authors and also corresponding authors

**Extended Abstract / Synopsis**

**With 1.85 billion tons produced per year, steel is the most important material class in terms of volume and environmental impact. While steel is a sustainability enabler, for instance through lightweight design, magnetic devices, and efficient turbines, its primary production is not. For 3000 years, iron has been reduced from ores using carbon. Today 2.1 tons $CO_2$ are produced per ton of steel, causing 30% of the global $CO_2$ emissions in the manufacturing sector, which translates to 6.5% of the global $CO_2$ emissions. These numbers qualify iron- and steel-making as the largest single industrial greenhouse gas emission source. The envisaged future industrial route to mitigate these $CO_2$ emissions targets green hydrogen as a reductant. Although this reaction has been studied for decades, its kinetics is not well understood, particularly during the wüstite reduction step which is dramatically slower than the hematite reduction. Many rate-limiting factors of this reaction are set by the micro- and nanostructure as well as the local chemistry of the ores. Their quantification allows knowledge-driven ore preparation and process optimization to make the hydrogen-based reduction of iron ores commercially viable, enabling the required massive $CO_2$ mitigation to ease global warming. Here, we report on a multi-scale structure and composition analysis of iron reduced from hematite with pure $H_2$, reaching down to near-atomic scale. The microstructure after reduction is an aggregate of nearly pure iron crystals, containing inherited and acquired pores and cracks. Crucial to the reduction kinetics, we observe the formation of several types of lattice defects that accelerate mass transport inbound (hydrogen) and outbound (oxygen) as well as several chemical impurities within the Fe in the form of oxide islands that were not reduced. With this study, we aim to open the perspective in the field of carbon-neutral iron production from macroscopic**

**processing towards a better understanding of the underlying microscopic reduction mechanisms and kinetics.**

**Introduction and motivation**

Steel is by far the most important metallic material, both in terms of quantity and the breadth of industrial applications, ranging from transportation, infrastructure, construction, machinery to safety. It is also an enabler of clean energy conversion technologies, such as wind or solar thermal. Steel can be recycled practically infinitely by collecting and re-melting scraps. When averaged over all steel grades and scrap types its global recycling rate is about 70%, making steel the world's most recycled material, and, in absolute tonnage, amounting to more than all other recycled materials combined[1]. However, due to the steel's role as a backbone material of economic development, scrap-based secondary synthesis alone cannot satisfy the global demand. Therefore, huge quantities of iron are produced each year by conventional primary synthesis, via reduction of iron ores in blast furnaces, using carbon as a reductant[2]. The current annual consumption of iron ores for this process amounts to a gigantic 2.6 billion tons, producing about 1.28 billion tons of pig iron through this route[3]. Each ton of steel produced by this conventional primary synthesis, i.e. through a blast furnace followed by a basic oxygen converter, creates about 2.1 tons of $CO_2$. Currently, more than 70% of the global iron production is made through this process route[2].

These numbers make steel one of the most staggering single sources of greenhouse gas on the planet, with about 6.5% of all $CO_2$ emissions, i.e. 30% of all $CO_2$ emissions from the manufacturing sector[4]. Global growth rate projections suggest a massive further increase of these emissions at least up to the year 2050 if no technology changes are implemented.

Beyond the underground $CO_2$ storage, a technique that might cause harmful effects on soil and water, three strategies are currently explored to mitigate this dramatic contribution of primary iron synthesis to global warming.

The first one is an increase in recycling. However, secondary production of steel still causes up to 1 ton $CO_2$ per ton of recycled steel, due to the use of graphite electrodes in the electric arc furnaces and the electrical power which is mostly generated from fossil fuels. Also, there is simply not enough scrap available to cover the total demand, due to the longevity of many steel products, for instance, in buildings and vehicles[5].

The second scenario is the use of electrolysis, similar to the primary production of aluminium via the Hault-Perrault process[6]. However, the iron oxide's high melting point and the chemically aggressive nature of the high-temperature liquid salts make this processing route currently less attractive for commercial production. Promising in this field, however, is the recent progress in low-temperature electrolysis, using ionic liquids as solvents[7–9].

The third, and currently most viable alternative, lies in the use of hydrogen as a reducing agent (instead of carbon), provided it comes from sustainable sources[10]. Hydrogen-based reduction schemes need to consider three thermodynamic constraints in the design of reactors. (1) The net energy balance for the complete reduction of iron oxide to iron with hydrogen is endothermic, i.e. it requires external energy to proceed[10]. (2) In many current transition techniques, hydrogen is not used as the only reductant, but it is mixed with carbon carriers in the same reactor[11]. This means that the catalytic splitting of the injected dihydrogen into reactive atomic hydrogen ($H_2 \leftrightarrow 2H$) and its reaction with the oxide directly competes with the Boudouard reaction ($C+CO_2 \leftrightarrow 2CO$) providing the reactive CO and the reaction of this molecule with the oxide[10]. Also, the reaction product, i.e. water, must be removed from the reaction zone as it can re-oxidize or block the reduction front. This is a challenging detail as water or, respectively, the oxygen, desorbs and diffuses only slowly compared with the reductant molecules. High partial pressure of water has indeed been shown to act detrimental to the nucleation and growth of iron on wüstite surfaces[13,14]. (3) The availability of green hydrogen (made from sustainable sources) is currently by far too small to mitigate the steel industry's greenhouse gas output. This means that the reduction via grey hydrogen has to serve as transition technology with a reduced mitigation effect[7,15]. Grey hydrogen, which makes more than 95% of the hydrogen market, comes from steam reforming and partial oxidation of methane, coal gasification and wet coke gas production.

**Approaches to hydrogen-based reduction of iron oxide and the role of microstructure**

The hydrogen-based reduction scenario with the currently highest technology-readiness level is to inject hydrogen into existing blast furnaces, in addition to coke[14,15]. The coke lends the furnace its permeable structure. This structure is needed to allow gas percolation as well as slag and metal outflux[10]. Current pilot operations use either grey hydrogen (from gas reforming) or coke gas. The latter is a mixture containing up to 65% hydrogen and more than 20% methane, available in integrated steel factories from the coke plant[17,18].

Yet, the underlying microscopic processes and some aspects associated with the basic thermodynamics and kinetics within the blast furnace are not well understood and can thus not be controlled or further tuned for higher reduction efficiency and improved $CO_2$ balance. A possible consequence of this is that the injected hydrogen partially evades without a substantial contribution to the reduction reaction and thus without markedly reducing $CO_2$ emissions, even though hydrogen diffuses and percolates much faster than CO and $CO_2$. Another efficiency limit is that the reaction is endothermic, hence, more carbon-carriers are required to keep the temperature high enough when increasing the hydrogen amount injected.

The second technology option is the direct reduction of solid iron ore pellets by hydrogen at temperatures above 570 °C[7,19,20]. In this case, the reduction occurs from hematite ($Fe_2O_3$) to magnetite ($Fe_3O_4$), and further to wüstite (FeO) and metallic iron (Fe).

Several macroscopic studies exist on the reduction of iron oxide exposed to different gas mixtures (in part including hydrogen) in the temperature range of 500-900 °C[21–24]. However, key aspects that influence the reduction kinetics in terms of the microstructure, nano-chemistry, porosity, and mechanics of the ores and partially reduced products are not well understood.

These unknown parameters and mechanisms have in common that they act on micro- and even near-atomic length scales. Thus, the next level of insight into the carbon-free reduction of iron ores requires direct observations at these small scales, probing both, structure and chemistry.

A better understanding of these processes would enable the design of suited ore pellets and process conditions with respect to ore preparation, pre-processing, enrichment strategies, microstructure, dispersion, porosity, grain size, texture, thermo-mechanical conditions, and chemical composition regarding the pellets, as well as reduction gas mixtures and reactor design.

**Experimental set-up and global kinetics of reduction of hematite with hydrogen**

For studying the influence of microstructure on the hydrogen-based reduction of iron ore pellets, we conducted a multiscale investigation of an isothermal reduction of commercial hematite direct reduction pellets (8 mm pellet size; 0.36 wt.% FeO, 1.06 wt.% $SiO_2$, 0.40 wt.% $Al_2O_3$, 0.73 wt.% CaO, 0.57 wt.% MgO, 0.19 wt.% $TiO_2$, 0.23 wt.% V, 0.10 wt.% Mn as well as traces of P, S, Na, K), exposed to pure hydrogen at 700°C in a static bed, using electron microscopy, electron backscatter diffraction, electron channelling contrast imaging and atom probe tomography.

Figure 1a shows the mass change of a hematite ore pellet during its reduction in pure dihydrogen at 700 °C under a constant gas flow rate of 30 L/h. The reduction degree is shown in Figure 1b. The reduction rate drops fast in the initial stage, slows down after approx. 2400 s and reaches completion after approx. 6000 s. More specifically, Figure 1c shows the reduction rate as a function of the fraction of reduced material, allowing a better separation of the kinetic stages in such gas-solid reactions[10,25]. Figure 1b and c summarise the different steps in the reduction process from $Fe_2O_3$ (hematite) to $Fe_3O_4$ (magnetite) to FeO (wüstite) and finally to α-Fe (α-iron). The dotted lines at the reduction degree of 0.11 and 0.33 indicate the stoichiometry pertaining to the complete reduction from $Fe_2O_3$ to $Fe_3O_4$ and $Fe_3O_4$ to FeO, respectively. The observed reduction sequence agrees with the previous literature on the hydrogen-based hematite reduction at temperatures above 570 °C[10].

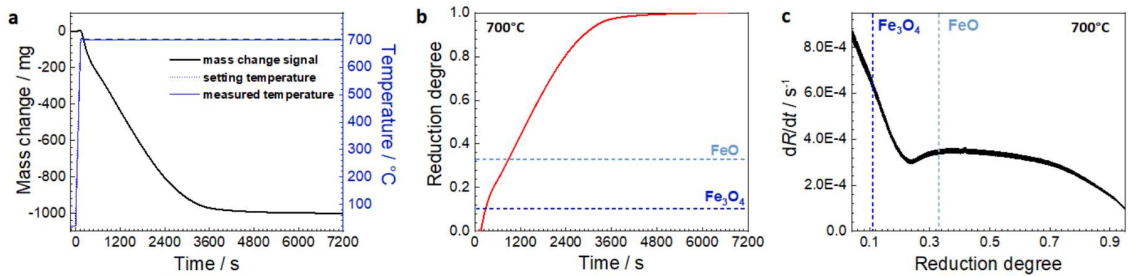

**Figure 1: Reduction kinetics of hematite ore reduced in hydrogen at 700 °C at a gas flow rate of 30 L/h. (a)** Mass change and temperature vs. time. **(b)** Reduction degree (normalized values from (a)) vs. time. The dotted lines at 0.11 and 0.33 are theoretical reduction degrees for a complete reduction from $Fe_2O_3$ to $Fe_3O_4$ and $Fe_3O_4$ to FeO, respectively. **(c)** Reduction rate vs. reduction degree. The end of the abrupt slope between 0.11 and 0.15 reduction degree marks the end of the first stage of the reduction. It is characterized by the fast transition from $Fe_2O_3$ to $Fe_3O_4$ and the subsequent stage of decelerating reduction rates ends at the local minimum between 0.28 and 0.40, indicating the $Fe_3O_4$ to FeO transition. The further reduction of the FeO into Fe becomes then the rate-limiting process.

The global kinetic data show that the reduction rate initially decreases very rapidly, going through a local minimum (at a reduction degree of ~0.23) and then enters into a sigmoidal shape. The first stage with its rapid decay ends between 11 and 15 % of the overall reduced material. As suggested by stoichiometry, this transition region marks the end of the first stage in reduction, namely, the reduction $Fe_2O_3 \rightarrow Fe_3O_4$. The subsequent stage, characterized by a steep steady slow-down of the reduction rate corresponds to the transition from $Fe_3O_4$ to FeO. The mass change at the end of the $Fe_3O_4$ to FeO transition overlaps with the onset of the FeO→Fe reaction. We assume, that at the local minimum, the mass loss due to the FeO→Fe transition is more dominant than that of the $Fe_3O_4$→FeO transition, after which the further reduction of the FeO becomes the prevalent process. The significant decrease in the reduction rate for the first two transition steps indicates that they are diffusion controlled. This suggests that the transport of reactants to or products away from the reaction interface are the rate-determining steps. This implies that the reduction rate will further decrease with the steady increase of the product layer thickness. It is likely that the transport of oxygen from the internal reduction front to the outer surface of the relevant reduced phase is rate determining as H diffuses inbound much faster than does O outbound.

The sigmoidally shaped wüstite reduction step, with a low reaction rate at the beginning and end, is supposedly a rather nucleation-controlled process, especially at the initial stage. This kinetic interpretation is plausible as the $Fe_2O_3$ to $Fe_3O_4$ reduction as well as the $Fe_3O_4$ to FeO reduction only stand for modest stoichiometric O losses of 1/9 and 1/4 unit O, respectively, whereas in the final reduction step, FeO loses a full unit of O. Nucleation limitation is less likely to be rate controlling for the reduction of $Fe_2O_3$ and $Fe_3O_4$ owing to rather high structural similarity[26–28]. In contrast, Fe has a very different crystallographic structure and volume than FeO, which makes its nucleation difficult. Thus, the compositional and structure fluctuations required for nucleation of Fe inside the FeO are larger than those for the preceding reduction steps. This means that in this last stage nucleation is rate-limiting, even though gas diffusion is comparatively fast, aided also by transformation-driven pore formation.

It should be noted that the plot of conversion rate against conversion degree for the wüstite reduction process (Figure 1c) does not exactly exhibit a sigmoidal shape, but it has a plateau region starting at around 0.3-0.45 reduction degree, Figure 1c. This feature indicates that the beginning of the reaction is determined by the Fe nucleation rate but the free expansion of these nuclei until impingement gets decelerated by a growth-limiting effect, which will be discussed later in more detail.

The FeO to Fe transition process can be divided into several kinetic steps: (1) $H_2$ gas molecules diffuse to the surface and (2) dissociatively adsorb there to form adsorbed H. (3) The adsorbed H diffuses inbound and reacts with O, which then diffuses to the surface from the internal FeO/Fe reaction front where equilibrium prevails. This drain of the reaction product O away from the reduction front leads (4) to Fe nucleation and (5) the steady release and supply of O at the reduction front from the oxide, its (initially slow) entropy-driven outbound diffusion and its consumption at the outer surface by reaction with H into water. These effects result in the gradual enrichment of O at the reaction front, creating a gradient of the chemical potential that drives the transport of O from the internal reaction front toward the outer surface. If steps (1), (2) or (3) were the slowest ones, then the whole reduction process should maintain the same kinetics, which is definitely not the case as evidenced by our data. Step (4) is a nucleation-controlled step as discussed above. Considering the extremely small size of H, the abundant H supply, and the correspondingly high H gradient from the surface to the reaction interface, it is plausible to assume that not the availability of H but the strongly retarded outbound solid-state diffusion of O, associated with the thickening of the Fe product layer surrounding the residual FeO, is the main rate-limiting step in the overall wüstite reduction.

Although for the $Fe_2O_3 \rightarrow Fe_3O_4$ and $Fe_3O_4 \rightarrow FeO$ transitions the respective $Fe_3O_4$ and FeO product layers also serve as barriers against the outward diffusion of O from the reaction interface to the surface, the final Fe product layer surrounding the wüstite suppresses mass transfer much more efficiently, owing to the very low diffusion coefficient particularly of O in the bcc Fe lattice. Also, both FeO and $Fe_3O_4$ contain multiple defects at elevated temperature[29]. Thus elemental diffusion through such constituents should be faster than through bulk bcc Fe. Yet, the equilibrium oxygen partial pressure for Fe/FeO ($3.0 \times 10^{-22}$ atm, calculated by using the HSC Chemistry 6.0 software) at 700 °C is several orders of magnitude lower than that for the pairs FeO/$Fe_3O_4$ ($1.1 \times 10^{-21}$ atm) and $Fe_3O_4$/$Fe_2O_3$ ($9.9 \times 10^{-12}$ atm). This renders the oxygen gradient, corresponding to the chemical driving force for diffusion from the reaction interface to the surface, significantly lower in Fe in comparison with that in FeO and $Fe_3O_4$, when assuming the same thickness.

This first analysis shows that a more detailed interpretation of the reduction kinetics requires information about the microstructure of the pellets, both in their initial oxidic state as well as in their partially reduced state(s). Also, more information about the distribution of the lattice defects, the pore evolution, and dense iron layers is needed to interpret the reaction kinetics.

Of special interest in this context are the later stages of the reaction, characterized by the reduction of the wüstite into iron. Figure 1b and c clearly reveal that the wüstite reduction rate decreases substantially during the last 20-30 % of the reduction. The lower mass transport of O through the Fe product layer is most likely the main factor for this slow kinetics. The fact that the reduction rate remains nearly constant or at least does not decrease much between reduction degrees of 0.3 to 0.7 are supposed to be due to iron nuclei growth, which is important at the beginning of the wüstite reduction. This leads to an initial increase in the reduction rate and to the formation of pores and cracks due to volume mismatch. Such additional open volume and new surface regions reduce the need for slow mass transport exclusively through the solid-state phase. The last 20-30% of the reaction, however, seems to be governed by an increasing dominance of bulk diffusion.

These findings are consistent with earlier reports stating that the hematite is reduced very rapidly, while the later stages, dominated by wüstite reduction, are more sluggish[30,31]. Microstructure observations can thus inform, which defects can enhance mass transport during this reaction stage.

**Structure and composition analysis at macroscopic and microscopic length scales**

When optically imaged, iron oxides can exhibit a variety of different colors, mainly depending on the oxidation state of the Fe. Figure 2a shows the porous iron ore pellets prior to the reduction. It has a distinct reddish color, typical for its hematite-rich composition. This color vanishes, turning black, when the iron is reduced to metallic Fe through the reduction with H, Figure 2b. The samples were cut and embedded (Figure 2c) for further light (Figure 2d) and electron optical investigation, Figure 2e and f.

Backscattered electron (BSE) imaging of the unreduced pellet in the scanning electron microscope (SEM), Figure 2e, reveals that the microstructure consists of grains with sizes of up to tens of μm. They are separated by a fine mesh of pores, ranging from nm to multiple μm in size. The influence of the single crystal effects and grain size of the ore on reduction kinetics has been studied in the literature[32,33], reporting for the case of magnetite, that the reduction kinetics gets faster with smaller grain size.

The phase map from electron backscatter diffraction (EBSD) in Figure 2f shows that the initial ore pellet is predominantly comprised of hematite ($Fe_2O_3$) making up 99.3% of the volume, and a minor amount of magnetite ($Fe_3O_4$) of 0.7%. Elemental maps obtained by energy-dispersive X-ray spectroscopy (EDX) in the SEM are displayed in Figure 2g for some abundant slag elements[34,35], i.e. Ti, Mg, Na, V. The elemental maps highlight the compositional complexity on this scale, with a relatively homogeneous Ti distribution, while localized enrichments in Mg, Na, and V appear in intergranular regions likely in the form of oxides, which is certainly related to the composition of the slag elements in the pellet and their solubility in hematite. The EBSD map in Figure 2h shows that after reduction (700 °C, 2 h) 94% ferrite and 6% of retained oxide were obtained. Although the background signals are high in the EDX maps, compared with the initial ore pellet, and at that scale, all gangue elements appear to be localized in the vicinity of residual oxide clusters. It is worth noting that all the metallic slag elements in the pellet, such as Ti, Mg, Na, and V, possess a significantly higher affinity for oxygen than Fe, and hence are markedly more difficult to be reduced. The quite homogeneous appearance of the slag elements alongside with Fe on this scale is probably due to the coarse resolution. Below, we study the oxides therefore at high resolution using atom probe tomography.

Figure 3 shows the porosity and cracks in the pellets in the initial and reduced states. While the initial sample is characterized by the pore structure inherited from pellet manufacturing (porosity of 28 vol.%), the intermediate reduction stages reveal the formation of a large number of additional lengthy and small cracks and pores in the microstructure. The reduction at 700 °C for 10 min results in an increase in the free volume by about 5% and a decrease in the average diameter of the pores from 1.70 to 0.41 µm. This strong decrease in the average diameter is due to the huge number of small pores with a size below 2 µm, Suppl. Fig. S1. After 2 h reduction at 700 °C the porosity increases by about 13%, and the average diameter changes from 0.41 to 0.86 µm. This observation means that the microstructure of the reduced pellets does not only inherit its porosity from preceding processing, but also acquires additional free volume, together with multiple lattice defects (e.g. dislocations and cracks) due to the volume mismatch and the associated stresses between the educt and product phases.

The presence of porosity in iron ore pellets has been discussed in the literature. The pellets consist of iron ore fines that either arise during the processing of high-grade iron ores (i.e. >60% Fe content), or during a beneficiation process in which most of the gangue minerals are removed from low-grade iron ores, thereby enriching their iron content. Considering the generally highly heterogeneous composition of the natural raw material combined with different processing steps that require the admixture of additives such as binders, fluxing agents, or carbon, it is not surprising that the resulting pelletized iron ores also inherit a complex pore structure across different length scales[36]. Besides this initial porosity state, it is also important to understand that the porosity changes during the course of the reduction. This is due to (a) the net volume loss as the oxygen gets removed; (b) the volume changes associated with the phase transformation sequence from hematite to wüstite and finally to iron; (c) locally trapped gas (e.g. the water) pressure aiding crack opening and (d) the micromechanical boundary conditions which create internal stresses that lead to delamination and decohesion at the hetero-interfaces[37,38]. The influence of the initial porosity on reduction kinetics was studied by Weiss et al.[39], reporting also a change in porosity with progressing reduction. It was found that the degree of the acquired porosity depends on the reduction boundary conditions, particularly on temperature. Higuchi and Heerema[40,41] suggested that the presence of large pores with sizes above 15 µm enhanced the reduction rate more efficiently than small pores below 15 µm. This finding might be related to the relevance of capillary effects in smaller pores and cracks, as well as connectivity.

High porosity and tortuosity, i.e. the degree of the percolating topology of the internal free surface regions, provide fast diffusion pathways for the reactants. The accessibility of the internal surfaces for the reduction gas substantially accelerates the global reduction kinetics. It also plays an essential role in taking up and removing the oxidation product, i.e. the water that is formed during the redox reaction. The tortuosity of the pores away from the pellet surface is normally quite poor, especially at the early reduction stages when pathways (e.g. cracks) to the external atmosphere have only rarely been formed. Consequently, it is necessary to understand the effect of the initially existing closed pores on the reduction kinetics of the most sluggish step i.e. the FeO to Fe transition.

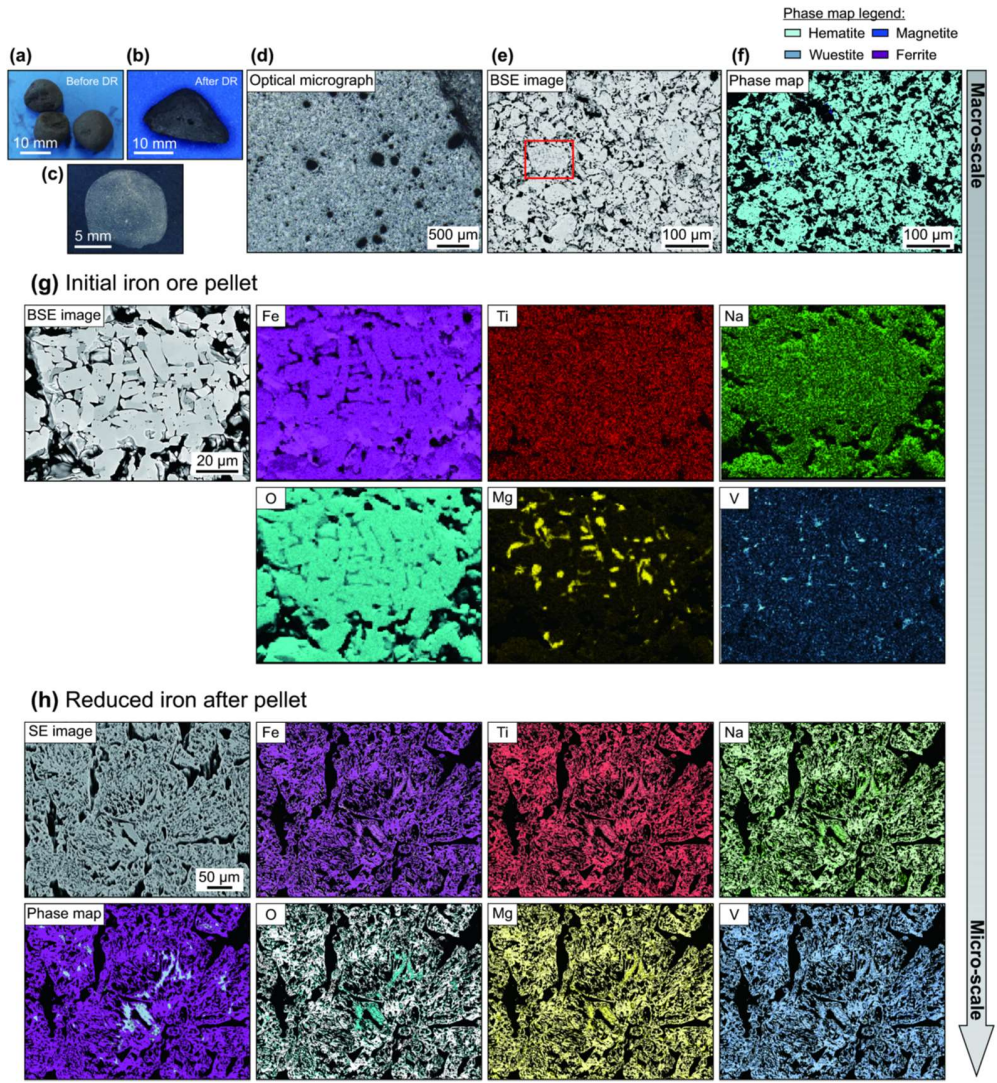

**Figure 2: Macro- and microscale structure and chemistry characterization of dihydrogen-reduced iron ore pellets. (a)** Porous iron ore pellets prior to reduction. (b) Iron ore pellet after reduction by using dihydrogen. **(c,d)** Optical images and micrographs of the as-reduced pellet, showing porosity. **(e)** Backscattered electron (BSE) imaging of the unreduced pellet in the scanning electron microscope (SEM). **(f)** Phase map obtained from electron backscatter diffraction (EBSD). **(g)** BSE image and elemental maps obtained by SEM energy-dispersive X-ray spectroscopy (EDX) for the initial pellet. **(h)** BSE image, phase map, and elemental maps obtained by SEM energy-dispersive X-ray spectroscopy (EDX) for the H-reduced pellet (700 °C, 2h).

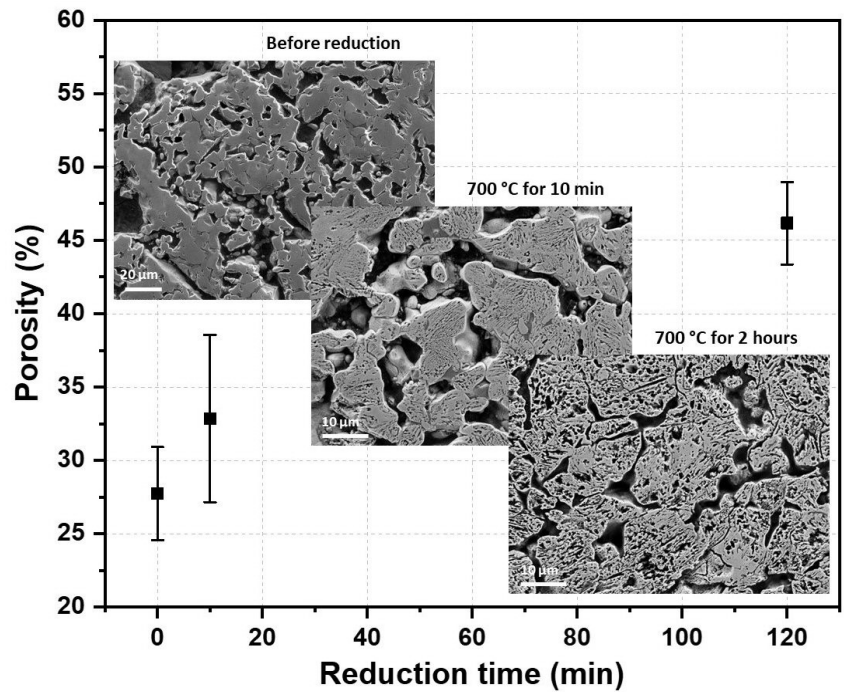

**Figure 3: Porosity analysis as a function of the reduction time.** Porosity prior to reduction, measured in sets of 2D metallographic images: 27.7±3.2 %; 700 °C for 10 min: 32.9±5.7 %; 700 °C for 2 hours: 46.2±2.8 %. Porosity analysis was conducted using the ImageJ software. For each condition, 12 SEM images were analysed. The porosity value was averaged over 12 measurements. It should be noted that the true porosity values in 3D might deviate from these values.

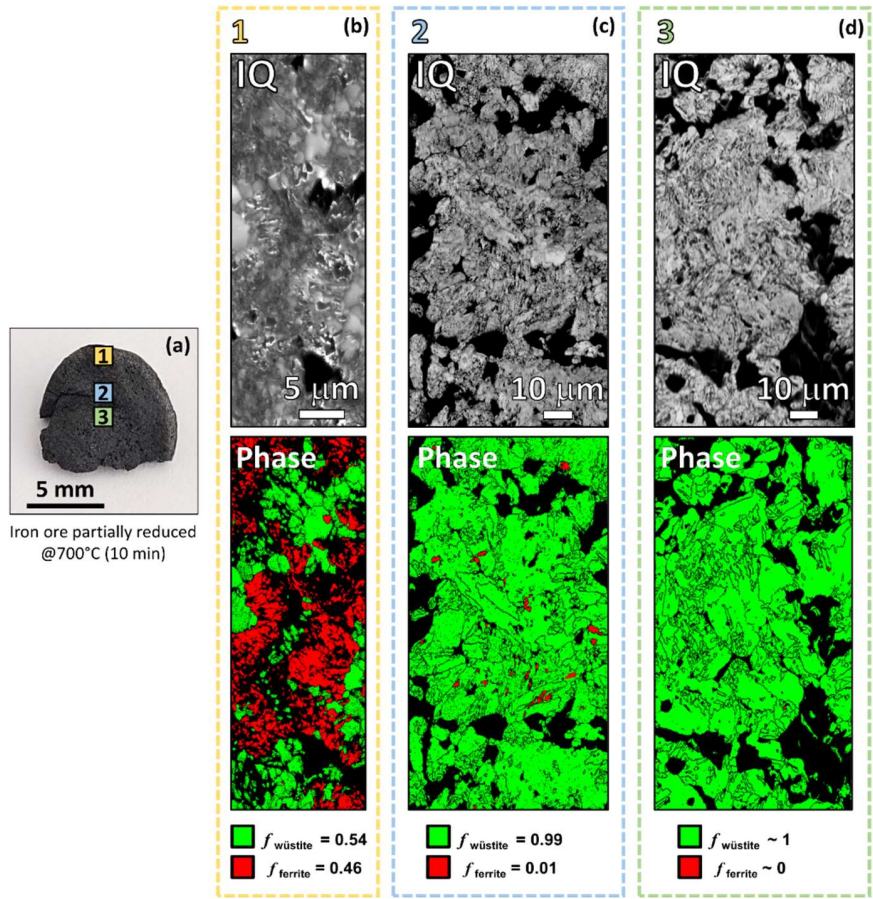

**Figure 4: (a)** Pellet partially reduced at 700 °C for 10 min under a $H_2$ flux of 30 L/h. The yellow, blue, and green frames labeled, respectively, as 1, 2, and 3 represent the regions analyzed by means of EBSD mapping. **(b)** Image quality (IQ) and corresponding phase map acquired from region 1 close to the specimen's surface. **(c)** IQ and phase maps from region 2, found between the surface and mid-pallet. **(d)** IQ and phase maps from region 3 at the central position of the pellet. In all phase maps, wüstite and ferrite are represented in green and red, respectively.

Figure 4 reveals that strong gradients in the reduction state exist (10 min at 700 °C). The images show the microstructure and the two dominant phases in different regions through the thickness of one of the pellets, as marked in Figure 4a. The outermost region of the pellet has been reduced into nearly 50% ferrite already after 10 min. (Figure 4b). This is attributed to a shorter solid-state diffusion distance from the surface to the reaction interface and thus a faster transport of O away from the interface. In the second microstructure image, taken closer to the center of the same ore particle, only 1% of the wüstite has been reduced to iron. In the center of the pellet, almost no reduction to Fe has taken place. Interestingly, most of the scattered Fe nuclei in Figure 4c are not formed close to the relatively large pores which obviously are part of the initial porosity. This means that the local atmosphere in the pre-existing closed pores at this stage is generally not reductive enough for Fe formation, due to the presence of the oxidation product, i.e. water.

At the transition stage from $Fe_3O_4$ to FeO, the oxygen partial pressure at the local atmosphere in the closed pores must assume a value between the $FeO/Fe_3O_4$ and the Fe/FeO equilibria, respectively. As the reduction progresses, the first nucleation of Fe appears, alongside the ongoing $Fe_3O_4$ to FeO transition, Figure 4c (note that $Fe_3O_4$ could not be resolved here). The closed pores cannot play an important role in mass transport, owing to lack of sufficient connectivity. Also, the equilibrium oxygen partial pressure for Fe/FeO is significantly lower than that for $FeO/Fe_3O_4$, which renders the local atmosphere for the closed pores unsupportive for the preferential formation of the Fe nuclei.

This means that any possible beneficial effect of the pre-existing closed pores on the reduction kinetics of the FeO to Fe transition becomes quite limited. Yet, closed pores possibly can shorten the diffusion distance through short-circuit paths, if the water formation and dissociation at the pore interface is comparatively fast.

Figure 5 provides a closer view of the microstructure of the partially reduced pellet. Almost all of the Fe nuclei are present at the grain boundaries, where nucleation of new phases and fast mass transport are often favored, Figure 5a. The microstructure images also reveal multiple delamination and crack events between ferrite and wüstite, due to the build-up of mechanical stresses associated with the transformations. More specific, the volume shrinkage from the initial α-hematite $Fe_2O_3$ to the magnetite spinel $Fe_3O_4$ is about 1.93%. The total volume difference between hematite and wüstite is even 19%. The final mismatch between FeO and Fe is large, about 44%[42]. This creates high mechanical loads among FeO and Fe which can be stored in the form of elastic energy, and/or is released in the form of plastic work or fracture events. Particularly the elastic work can act as an obstacle against Fe nucleation. Once a Fe nucleus is formed, the large volume change and the associated deformation cannot be accommodated by elastic distortion alone but leads to the formation of cracks, creep pores, and dislocations. This is revealed by the microstructure in Figure 5: the secondary electron image shows a variety of pores, cracks, and delamination features of different size and morphology. Figure 5d shows several such delamination effects between the ferrite and the adjacent wüstite. Local cracking and delamination provides new free volume in the system and thus contributes to locally enhanced kinetics of the wüstite reduction, as it enables faster diffusion.

Figure 5e-h show that the volume mismatch and the associated high-temperature deformation of the two adjacent phases lead to the substantial accumulation of dislocations, as revealed in Figure 5f through electron channeling contrast imaging (ECCI). The associated curvature of the lattice close to the interfaces is revealed in Figure 5g and h in terms of the gradual crystallographic rotation near some of the hetero-interfaces, a feature which translates to arrays of geometrically necessary dislocations. Transport of H (inbound) and O (outbound) through these crystal defects

might play an vital role in the kinetics, especially during the later stage of wüstite reduction when a relatively thick reduced iron layer surrounds the wüstite in a core-shell type morphology.

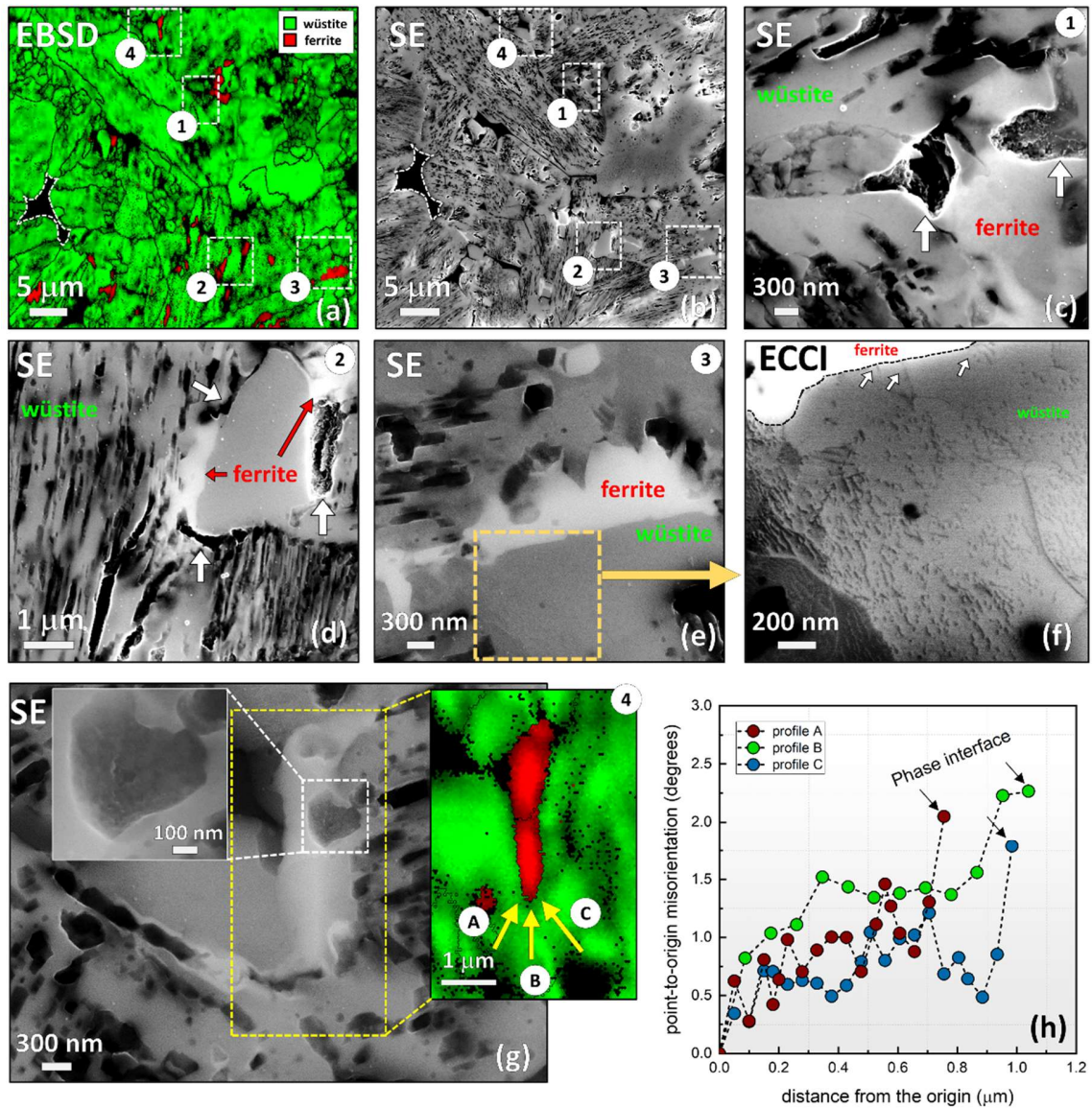

**Figure 5: Correlative EBSD analysis, SE, and ECCI for region 2 of the partially reduced pellet (10 min, 700 °C) in Figure 5. (a)** EBSD phase map where wüstite and ferrite are represented in green and red, respectively. **(b)** corresponding SE image of the map displayed in (a). The white frames named as 1, 2, and 3 highlight specific regions whose enlarged views are given in **(c)**, **(d)**, and **(e)**, respectively. **(f)** ECCI image obtained from the area delimited by the yellow frame in (e). **(g)** SE image and EBSD phase map from the area labeled as 4 in (a). The arrows A, B, and C were used as a reference to obtain local orientation gradients. **(h)** Point-to-origin local misorientation profiles acquired along the arrows A, B, and C in (g).

**Structure and composition analysis at near-atomic scale**

Some studies on H-based hematite reduction have discussed the important role of the oxide gangue[43,44], however, no chemical analysis at high spatial resolution has been conducted so far, and reports on the global chemical composition of ores do not provide a consistent picture on the effects of chemical impurities on the reduction kinetics and on the associated mechanisms.

Elemental inhomogeneity is not only a feature visible at the macro- and/or micro-scale, such as shown in Figure 2, but can also be observed at the reaction front at the atomic scale.

For revealing the possible role of non-ferrous oxides on the reduction kinetics at the reaction front, we studied the distribution of impurities by atom probe tomography (APT), before and after complete reduction. Details on the experimental protocols are given in the supplemental Figures S2-S5.

The three-dimensional atom maps obtained from the as-received hematite ore is shown in Figure 6a, in which Fe atoms are shown in pink and O atoms in cyan. The data set has an average Fe and O content of 44.4 and 51.6 at. %, respectively, translating to a stoichiometry ratio of 0.86 (see Table S2 for atomic composition details) although the expected stoichiometric ratio of hematite is 0.67. The O depletion in the composition can be explained by the loss of neutral O during field evaporation as desorbed neutral O atoms and also $O_2$ molecules do not contribute to the detector signal[45–47]. Figure S6 shows a multiple-event correlation histogram that reveals significant tracks indicative for the formation of neutral oxygen species through the decomposition of $FeO_2^+ \rightarrow Fe^+ + O_2^0$ and $FeO_3^+ \rightarrow FeO^+ + O_2^0$ during the APT measurement

The most abundant gangue elements such as Mg, Al, Ti, and V exhibit a homogeneous distribution in the initial iron ore on the nano-scale, as revealed by a nearest-neighbor analysis presented in Figure S8. Figure 6b shows a reconstructed 3D atom map of the reduced iron ore exhibiting a Fe content of 98.6 at.%, with a residual oxygen content of 0.01 at.%. Retained crystallographic information from the APT data confirms the body-centered cubic crystal structure (Figure S9). Major impurity elements such as Mg, Al, and Ti are not detected within reduced α-Fe but trace amounts of V at 100 ppm level are observed (see Table S2).

Another dataset displayed in Figure 6 contains a nano-sized oxygen-rich platelet embedded within the reduced α-Fe matrix, which appears to be crystallographically aligned along the {002} planes of the bcc structure (details in Figure S10). The composition of the remaining, trapped oxide is complex: it contains 33.3 at.% O and the rest consists of several metals 36.7 at% Fe and

19.1 at% V, as well as low amounts, below 3 at%, of the other gangue elements Mg, Al, and Ti, as revealed by the one-dimensional composition profile in Figure 6d. The nano-sized oxygen islands remaining also after reduction, originate from the gangue oxides and not from Fe-based oxide as they have low oxide-formation free energies at the reduction conditions according to Ellingham's diagram[48].

During the reduction process, oxidized species of less noble elements such as Al, Mg, and Na cannot be reduced. When they are relatively homogeneously distributed inside the iron oxide, they can either remain in the form of tiny oxide nano-particles, which might be thermodynamically not favorable (as the interface energy between oxide and metal is generally high), or they can be expelled from the reduction front into the remaining oxide. As the size of the remaining oxide shrinks, this oxide will become enriched by these species, and the highest concentration of the slowest diffusing and strongest bonded elements will accumulate at the interfaces. The enrichment of the less noble elements at the interface might significantly influence the further reduction.

The APT results also show that several gangue-related impurities indeed get captured in nano-sized metastable oxides, possibly on the way to the spinel phase formation. The observed ratio of O to metal ions is unlike reported for any typical Fe-rich oxide, including those in the Fe-V-O system[49]. A $Fe_2O$ oxide mineral was reported to be stable at high pressures such as those encountered in the Earth's core[50,51], and off-stoichiometric amorphous Fe-rich oxides with similar compositions have been reported in tribological layers [52]. Here, while embedded within the metallic lattice and in the harsh conditions of the reduction (i.e. 30 L/h of $H_2$ gas flow at 700 ºC), out-of-equilibrium, transient states are likely to appear.

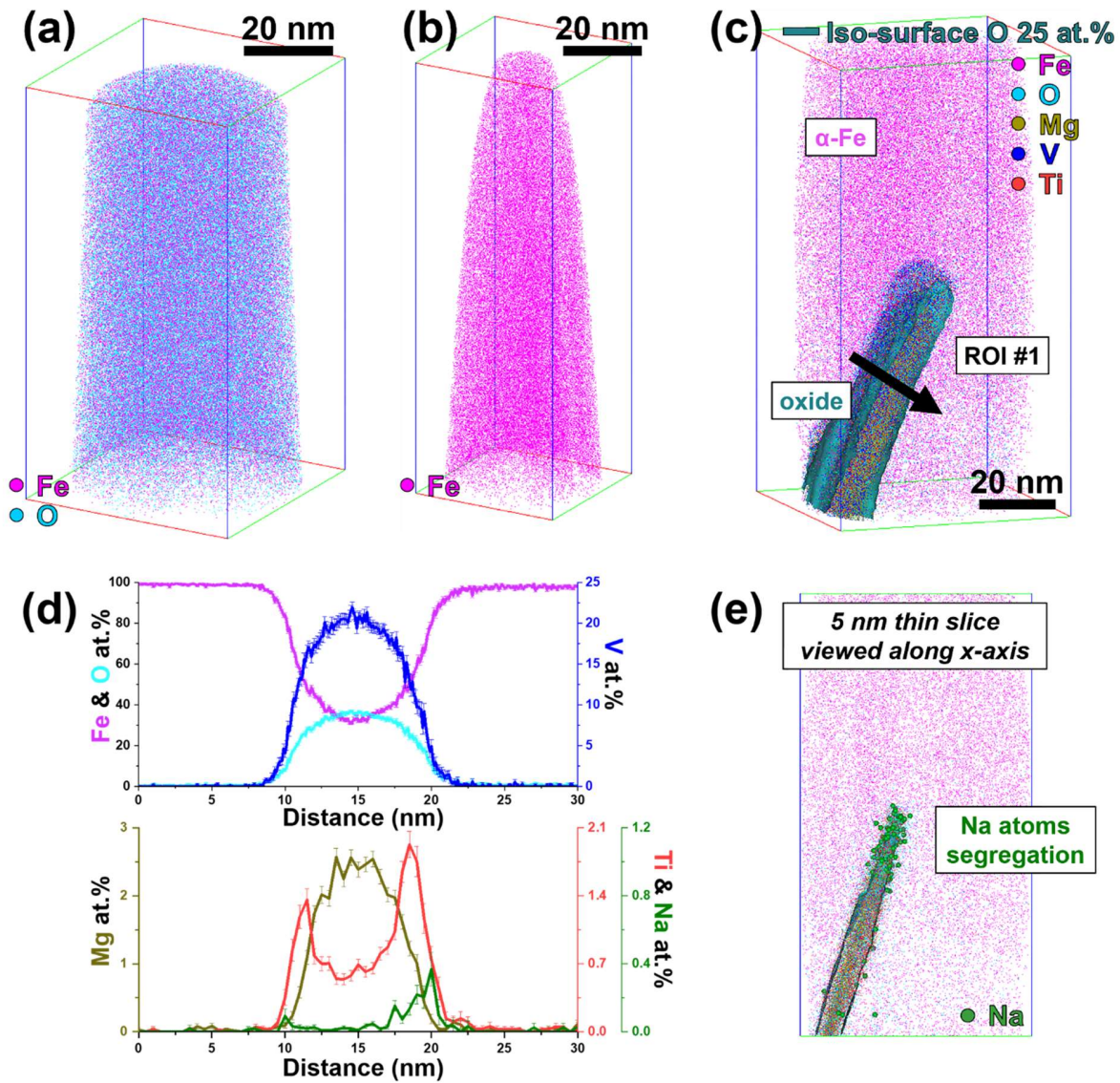

**Figure 6: 3D atom maps of (a) before and (b) after the direct reduction of as-received hematite ore.** Pink and cyan dots in the 3D reconstructions represent individual Fe and O atoms, respectively. **(c)** A reconstructed 3D atom map acquired from as-reduced ore. Iso-surface of O at 25 at.% (cyan) represents the interface between α-Fe and the wüstite oxide. The inset is a 3D atom map of a 5 nm thin slice. **(d)** Corresponding one-dimensional composition profile from the cylindrical region of interest, marked as ROI #1, (⌀15 x 30 nm$^3$) across the oxide nano-feature. **(e)** Na segregation at the interface between the retained oxide and the reduced matrix. Pink, cyan, golden brown, red, blue, and green dots represent the reconstructed positions of Fe, O, Mg, Ti, V, and Na atoms, respectively.

Some details of the composition of these remaining oxide particles are worth further exploring. It is conceivable that the harsh reaction conditions have enabled diffusion of these elements out of the Fe($O_x$) matrix, resulting in chemical partitioning across the metal/oxide interfaces and within the remaining oxide. The Na appears segregated at the interface. As the mobility of Na is comparatively poor due to its large ionic radius, Na is supposed to be one of the last gangue elements reaching the nano-sized oxides. The solubility of Na in the residual oxides is also expected to be low on account of the large difference in the ionic radii, which, accompanied by the belated arrival of Na, leads to its enrichment at the metal/oxide interface. From the tomogram of the oxide, most Na atoms (in green) are segregated at the apex of the oxide platelet, possibly a region of higher tensile stress where the large Na ions would be better accommodated.

Figure 7 reveals that at the later reduction stage FeO might be reduced to Fe via transient states due to the influence of certain gangue elements. Actually, this process can start much earlier than expected, as can be seen in Figure S11 of an impurity-doped oxide with Fe to O ratio of 3.3 just after 30 min reduction. A deeper reduction of the nano-sized residual oxide should lead to the formation of a remaining Fe-depleted oxide, as clearly illustrated in Figure 7 where phase separation between Fe-rich and Fe-depleted oxides appears. Ti and Mg are enriched in different oxides, as revealed in Figure 7a. Here the Fe to O ratio in the Fe-rich oxide is remarkably lower than that of its counterpart in Figure 6 (Table S2), which further indicates a higher reduction degree for the two-phase oxide.

Figure 7b shows a series of slices through the tomogram evidencing the separation between the two oxide variants. Figure 7c is a composition profile evidencing that Mg partitions to the Fe-rich oxide and Ti in the Fe-depleted oxide, while Na is segregated at the interface between the two oxides.

The Fe-rich oxide is mainly composed of V and Fe, with the V to Fe ratio remaining at a rather constant value slightly above 2 along the distance, which can be best ascribed to Mg-doped Fe$V_2O_4$ (also doped with other gangue elements, as documented in more detail in Table S2). The Fe-depleted oxide is mainly consisting of V and Ti, probably corresponding to $(V,Ti)_2O_3$, on account that the metal to O ratio of the Fe-depleted oxide (0.83) is quite similar to the Fe to O ratio of $Fe_2O_3$ (0.86) acquired by APT. Also, the solubility of Ti in $V_2O_3$ is relatively high. The identification of the two-phase oxide is reasonable also from a thermodynamic perspective, as the thermodynamic onset of the oxygen partial pressure follows the sequence: $V_2O_3 < FeV_2O_4 <$ FeO.

It needs to be emphasized that no matter which kind of Fe-containing oxide of gangue elements is formed during the wüstite reduction process, the equilibrium oxygen partial pressure for Fe/transient-state oxide must be at a value below that for Fe/FeO (e.g., the thermodynamic onset of Fe$V_2O_4$ formation is three orders of magnitude lower than that for FeO at 700 °C). This means that the oxygen gradient from the reaction interface to the surface will be lowered due to the formation of the transient-state oxide supposing the same thickness and thus the diffusion of O through the Fe product layer will be further reduced. The formation of the Fe-containing intermediate oxide plays a critical role in the significantly depressed kinetics during the final 20% reduction regime.

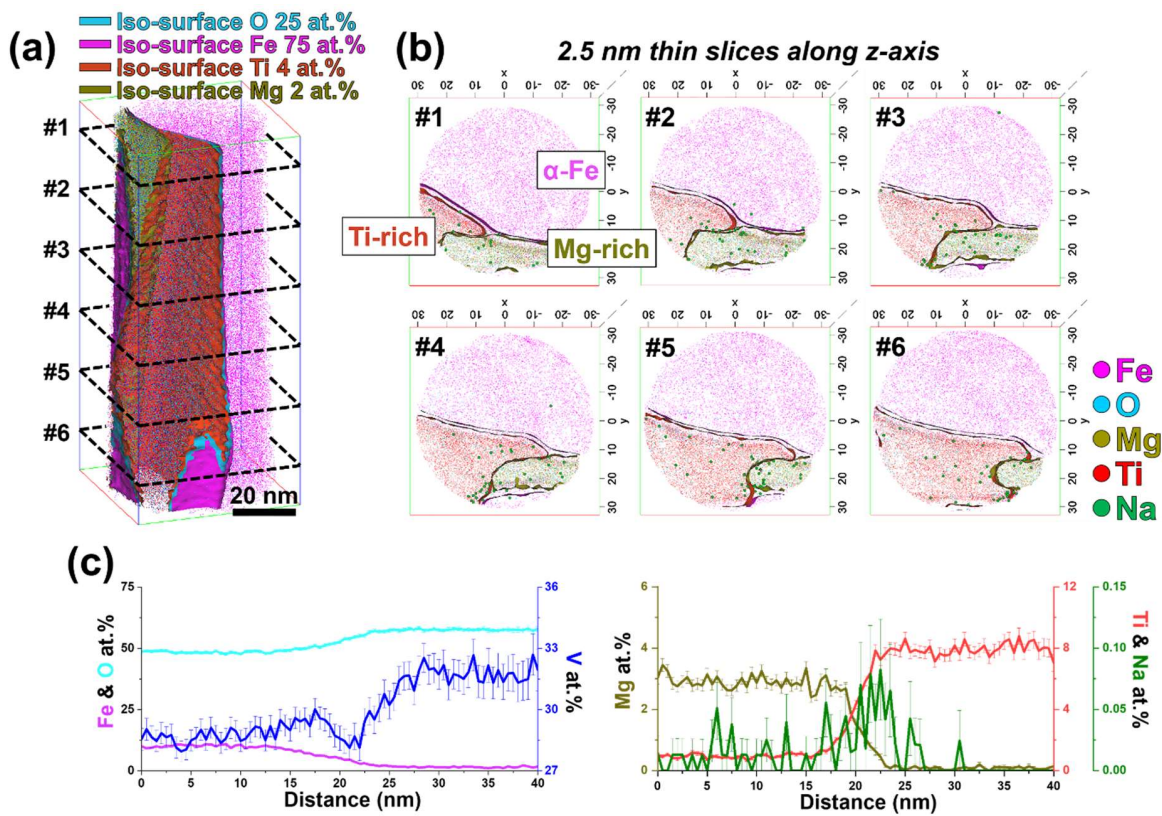

**Figure 7: Atom probe tomography analysis of partially reduced iron oxide. (a)** 3D atom map showing the distribution of selected major species with Mg- and Ti-rich oxides phases highlighted in 2 and 4 at.% iso-surfaces, respectively. **(b)** series of 2.5nm-thick slices through the reconstructed point cloud. **(c)** one-dimensional composition profiles (⌀7.5 x 40 nm³) across the adjacent oxides.

Understanding the behavior of gangue-related tramp elements allows us to identify kinetically relevant factors with respect to beneficial (catalysis effects) or harmful (delayed reaction rate) influence on the reduction reaction. Typical gangue compounds observed in direct reduction pellets are mostly Si-, Al-, Mg-, and Ca- containing oxides. While reports by Wang and Sohn[53,54] indicated that Ca- and Si oxides may alter the volume changes during reduction and affect the reduction rate, other papers found no clear trend of the influence of these oxides on the kinetics[21,55]. Depending on the content and conditions, MgO was reported to either lower the reducibility of sintered hematite[56] or promoting its reduction[43,57]. $Al_2O_3$ was observed to affect the reduction rate of magnetite as it triggers the formation of a network-like wüstite structure. The highest reduction rate was observed when adding 3 wt.% $Al_2O_3$ into the magnetite[58]. A $TiO_2$ content >0.5 wt.% was reported to significantly increase the reduction induced degradation, resulting in heavily cracked pellets[59]. The thickness of the formed Fe crystals gradually increases with increasing $V_2O_5$ additions in magnetite, which facilitates the aggregation and diffusion of Fe atoms[60].

Inconsistencies in some of these trends can now be better understood in terms of the fact that these studies were performed on the macro- and/or micro-scale and not at the near-atomic-scale at which the actual rate-limiting processes take place. Furthermore, revealing the effect of the gangue elements on the most sluggish step, i.e. the wüstite reduction, should be more beneficial for an in-depth understanding of the reduction kinetics, which, however, is rarely reported. Our APT results indicate that among the many gangue elements, special attention should be placed on those that can form mixed oxides with Fe and those that can be enriched at the Fe/oxide interface at the wüstite reduction stage.

The effects of the nano-scale chemistry on the reduction must be assessed further, as indeed most of the oxides encountered here play a critical role in the reduction process itself. There are many unknowns in the atomic-scale processes of how the local composition of the material affects the solid-state diffusion of elements, especially O and how the gangue elements accumulated at the metal/oxide interface influence the reduction kinetics. Of high relevance is also the effect of certain gangue elements on the formation of the transient-state oxide during the FeO to Fe transition and the interplay between the transient-state oxide and iron phases.

**Conclusions and outlook**

The impact of steel production via the conventional blast furnace and converter route, using carbon as a reductant, on the greenhouse gas emission is staggering, causing about 6.5% of all global $CO_2$ emissions. Reducing iron ores with hydrogen, via the solid-state direct reduction route, offers a viable alternative. The main challenge to solve, however, is to understand the origin of the sluggish reduction kinetics, particularly during the later stages of the wüstite-to-iron reduction.

Here, we conducted a study about the role of the micro- and nanostructure of the ores, both inherited and acquired during reduction, and about the nano-chemistry.

We find that the relatively easy nucleation of magnetite and the subsequent wüstite as well as the fast diffusion of O through the iron oxide product layer are the main reasons for the fast reduction kinetics of the hematite to wüstite. The third step, viz. the reduction of wüstite into iron is – during the final 20% reduction regime – nearly an order of magnitude slower. This effect is attributed to the slow Fe nucleation rate in the wüstite (relative to that required for the two preceding reduction steps) and to the sluggish mass transport (particularly of the outbound O) through the already formed iron product layers surrounding the wüstite.

The role of the pre-existing closed pores in the kinetics of the wüstite reduction is quite limited in view of the high nucleation barrier for Fe to form. Fe nuclei tend to preferentially form at the wüstite grain boundaries rather than at pores. This observation suggests that the pores are mostly closed rather than interconnected. The fast reduction at the external regions of the pellets is thus due to the short diffusion distances of O from the reaction interface to the surface. This supports the general observation that the outbound diffusion of O plays the dominant important role in the reduction kinetics.

The high volume mismatch between the wüstite and Fe leads to delamination cracks and high dislocation densities which might locally aid mass transport and thus the reduction process. This effect is, however, less relevant to notably mitigate the overall kinetic deceleration observed towards the later reduction stages.

Near-atomic scale chemical probing revealed the presence of nano-sized Fe-containing transient-state oxides and the accumulation of certain gangue elements i.e. Ti and Na at the metal/oxide interface, both of which are assumed to be another cause for the slow reduction kinetics at the late stages of the wüstite reduction.

Approaches to enhance reduction kinetics are, therefore, a high degree of plastic deformation and micro-cracking of the ore pellets (before and during reduction), providing more percolating free volume for accelerated mass transport; elimination of certain stable oxide-forming tramp elements stemming from the gangue; elimination of certain gangue elements that can be formed as ions with large radii; and smaller pellet sizes with – inside the pellets - smaller grain sizes with random crystallographic textures. A way towards enhanced reaction kinetics might thus also lead through an otherwise counterintuitive approach, namely, to make the pellets mechanically weaker (rather than stronger) so that they break in the reactor under well-defined loads during the wüstite reduction step, yet, without leading to undesired sticking.

**Acknowledgements**

We thank Uwe Tezins, Christian Broß, and Andreas Sturm for their support to the FIB & APT facilities at MPIE. We are grateful for the financial support from the BMBF via the project UGSLIT and the Max-Planck Gesellschaft via the Laplace project. BG and SHK acknowledge financial support from the ERC-CoG-SHINE-771602. KS is grateful to the IMPRS-SURMAT for funding of his scholarship.

**Data availability**

The data that support the findings of this study are available from the corresponding author upon reasonable request.

**Contributions**

## Supplementary Information

**Methods**

**Hydrogen direct reduction.** A Huasco hematite ($Fe_2O_3$) ore pellet for direct reduction with a diameter of 8 mm was placed on a quartz basket in an atmosphere-controlled thermogravimetry set-up with accurate temperature control. Then, the basket was suspended on the quartz hook of the thermobalance. Pure hydrogen gas with a flow rate of 30 L h$^{-1}$ was fed into the furnace from the bottom, passed the pellet, and flowed out at the top. After flooding the furnace with pure hydrogen, the pellet was heated with a heating rate of 5 Ksec$^{-1}$ and then isothermally heated at 700 °C for 2 h. All experimental probing parameters including mass change and temperature were recorded automatically. After the mass loss signals during the heating process reached an equilibrium, the ore pellet was left in the furnace to cool down to room temperature.

**Porosity analysis.** The porosity of iron ore and reduced products was analyzed based on the scanning electron micrographs at the magnification of 2000X (for the sample before reduction) and 5000X (for the samples reduced at 700 °C for 10 min and 2 h). The images were processed by the ImageJ software. The intensity threshold was carefully adjusted to segment pores from solid. In total, 12 SEM images were analyzed for each condition. The porosity value was averaged over 12 measurements. The equivalent diameter of each pore was calculated assuming a round shape of the pore: $d_{eq} = 2 \times \sqrt{A/\pi}$, where $A$ is the cross-sectional area of each pore. The distribution of the equivalent diameter of pores is shown in Figure S1 and the average diameters of the pores are summarized in Table S1.

**Atom probe tomography analysis.** The as-received hematite pellet and the as-reduced iron were prepared into atom probe specimens using a focused-ion beam (FIB). Needle-like shape specimens were fabricated following the sample preparation for powder samples according to Choi et al. (see Figure S2 for details) [1]. The as-prepared specimens were then mounted in a CAMECA LEAP 5000 XS system for APT analysis. A straight-flight-path atom probe 5000 XS system was used to collect ions with high detection efficiency (~80%) and to track any signs of molecular dissociation events [2]. A laser pulse energy of 40 pJ, a pulse frequency of 200 kHz, and a detection rate of 2% were set at a specimen temperature of 50K. Acquired data-sets were then analyzed using the Imago visualization and analysis system (IVAS) 3.8.4. All data-set was reconstructed using the standard voltage protocol [3] and saved as the *pos* format from which extended *pos* (*epos*) file was extracted for correlation histogram studies using MATLAB. Information on the voltage curve and detector event histogram for each data-set are presented in Figure S3-S5.

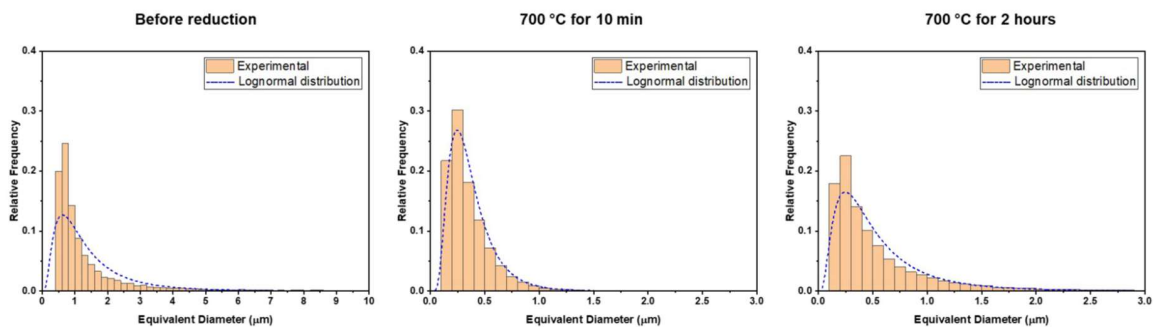

**Figure S1.** Distribution of the equivalent diameter of pores.

**Table S1.** The average diameter of the pores was calculated by averaging the mean size over 12 measurements.

| Condition | Average diameter (μm) |
|---|---|
| Before reduction | 1.70±0.08 |
| 700 °C for 10 min | 0.41±0.03 |
| 700 °C for 2 hours | 0.86±0.11 |

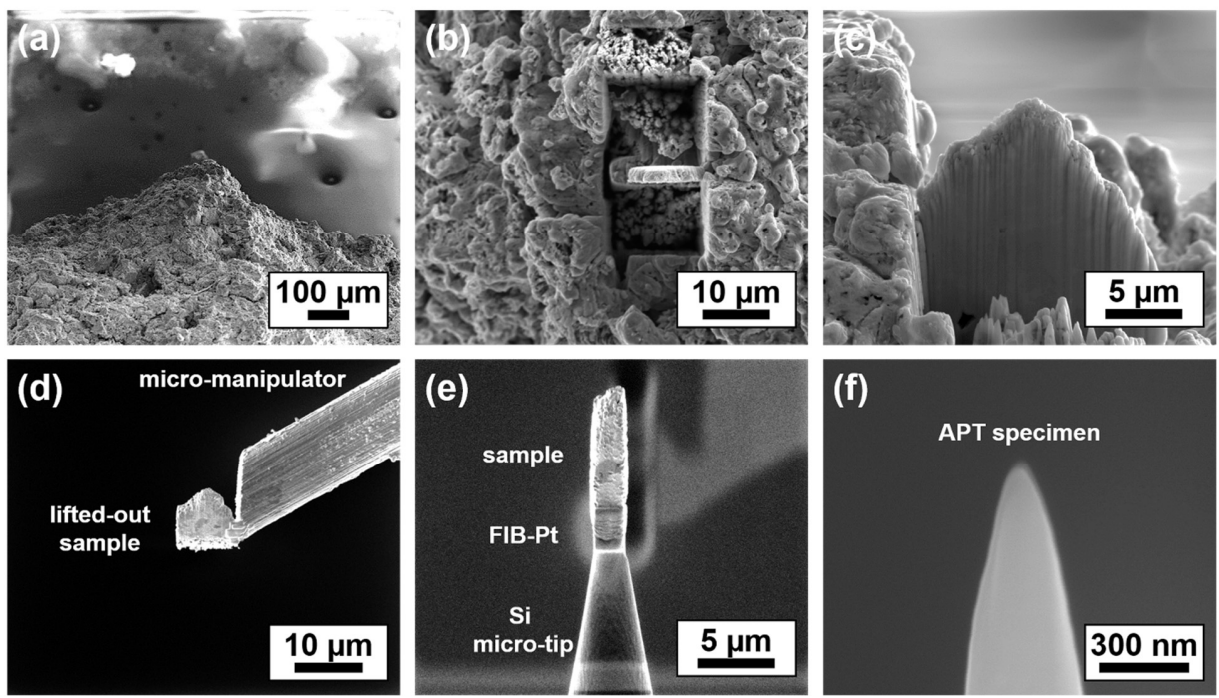

**Figure S2.** (a) 52° tilted e-beam image of a selected particle. APT specimen preparation process from as-received ore pellet: (b) Trenches were milled 15 µm in depths on the front and the back sides of the interested region (width ~2 µm). (c) The l-shape horizontal cut was made at the bottom of the sample in 52° relative to the ion beam column. (d) The sample was lifted out using an in-situ micro-manipulator. (e) Subsequently, the sample was welded with a FIB-Pt deposition on a Si micro-post only on the side where 52° cut was done. (f) Annular milling patterns from the top and decreasing diameters until the apex radius was below 100 nm with no pore. For Ga-damage cleaning process, all APT specimens were milled at low 5 kV and 8 pA for 60 s.

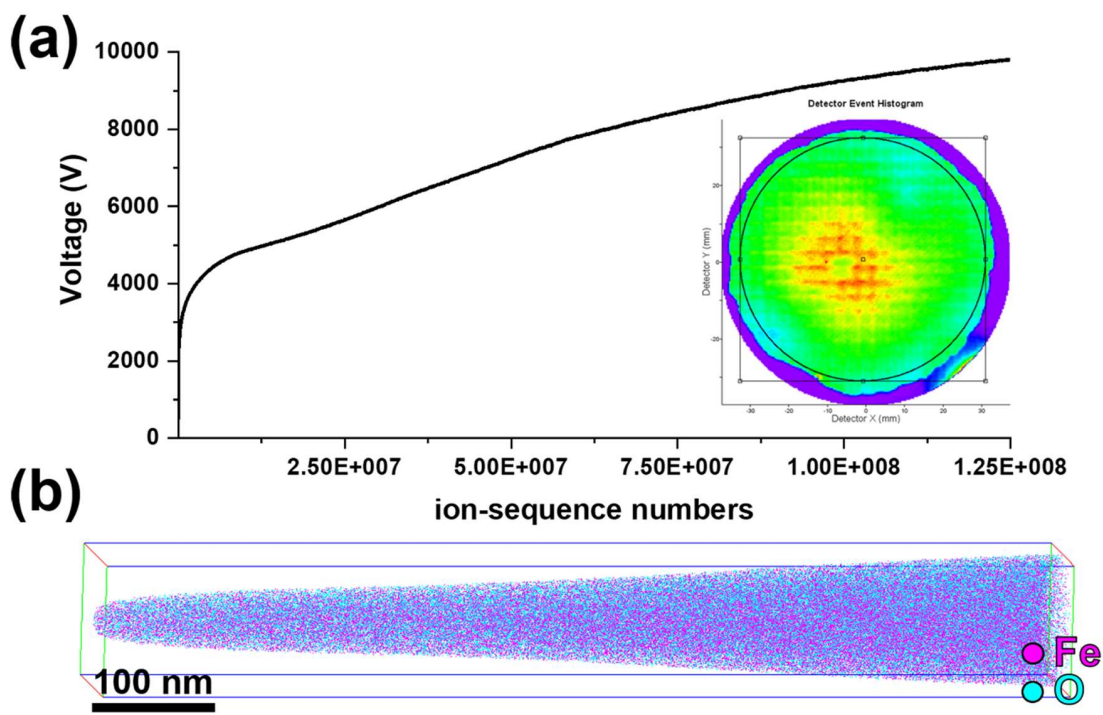

**Figure S3.** (a) Overall voltage curve and detector event histogram of as-received hematite APT data-set and (b) overall 3D reconstructed atom map. The reconstruction atom map of the first 60 million ions is presented in Figure 6(a). Average background level: 7 ppm ms$^{-1}$.

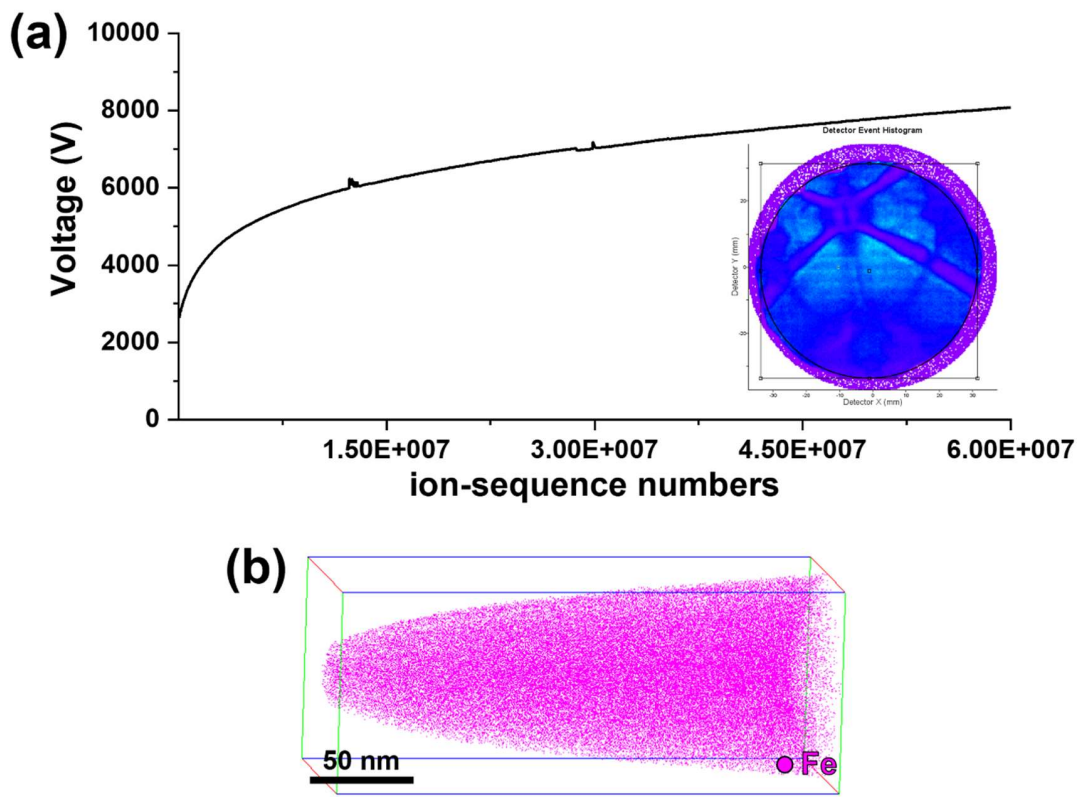

**Figure S4.** (a) Overall voltage curve and detector event histogram of as-reduced iron APT dataset and (b) overall 3D reconstructed atom map. The reconstruction atom map of the first 60 million ions is presented in Figure 6(b). Average background level: 5 ppm ms$^{-1}$.

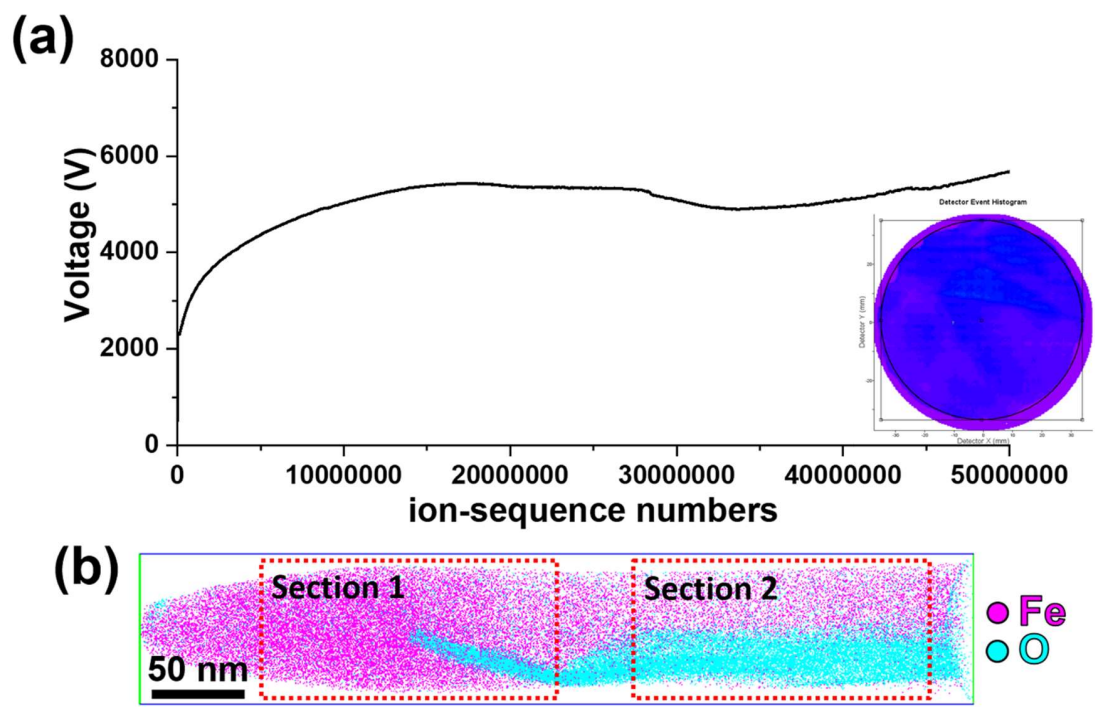

**Figure S5.** (a) Overall voltage curve and detector event histogram of the as-reduced iron APT data-set and (b) overall 3D reconstructed atom map. The reconstruction atom maps of the first 25 million ions (section 1) and the latter 25 million ions (section 2) are presented in Figure 6(c) and Figure 7(a), respectively. Average background level: 8 ppm ms$^{-1}$.

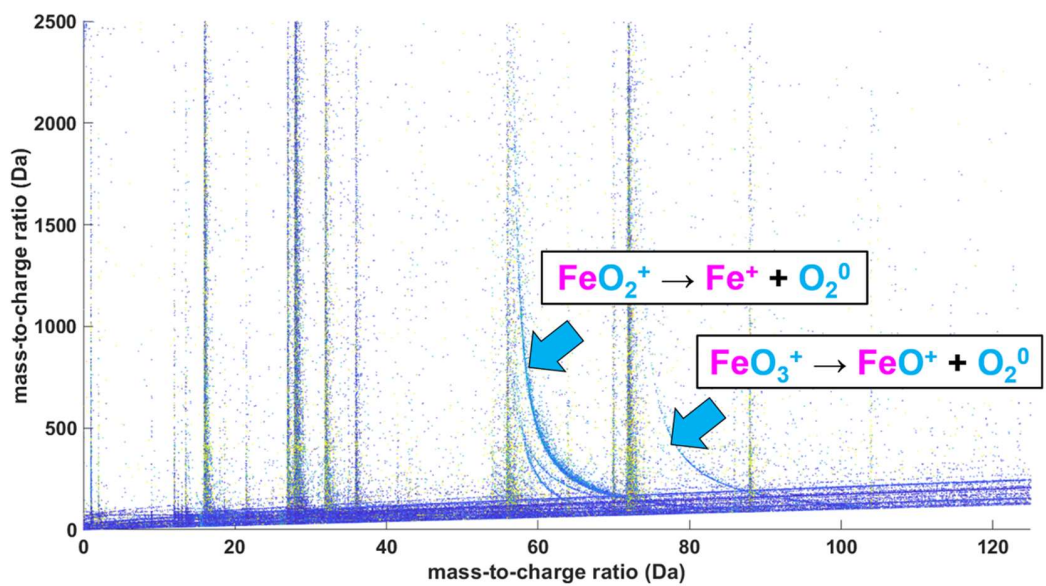

**Figure S6.** Multiple-event correction histogram for as-received hematite field evaporation acquired from Figure 6(a). Note that there are significant neutral formation trails of O and $O_2$ from dissociations of $Fe_xO_y$ ions. For comparison, the histogram of the as-reduced iron is shown in Figure S7.

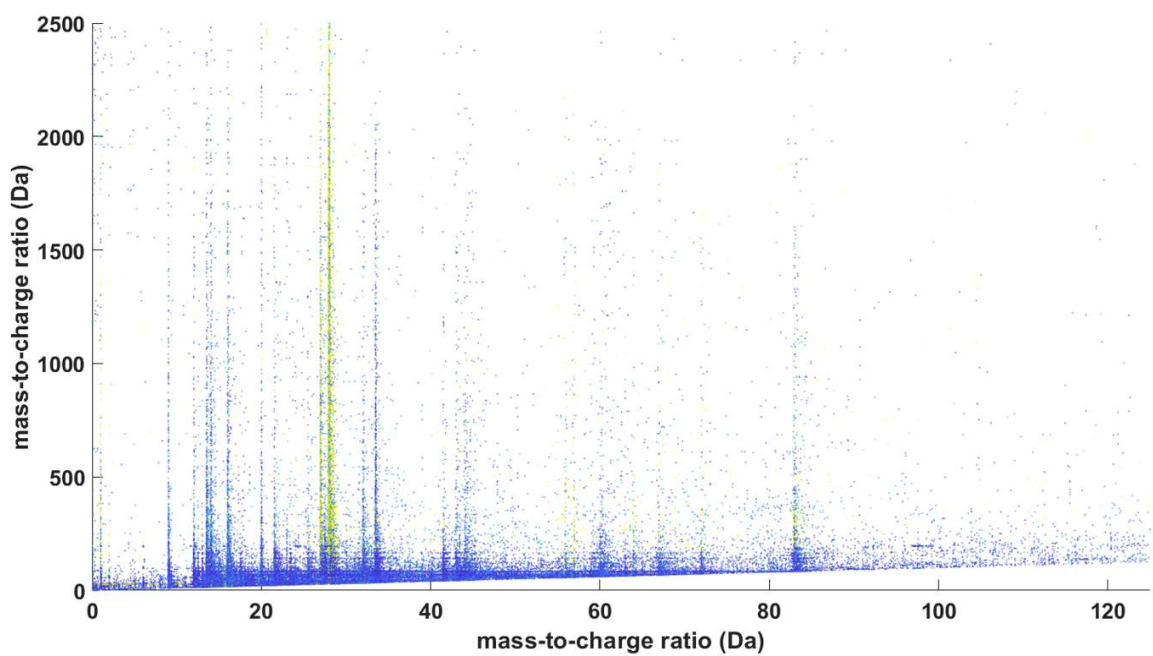

**Figure S7.** Multiple-event correction histogram for as-reduced iron field evaporation acquired from Figure 6(b). Note that there is no significant neutral formation trail.

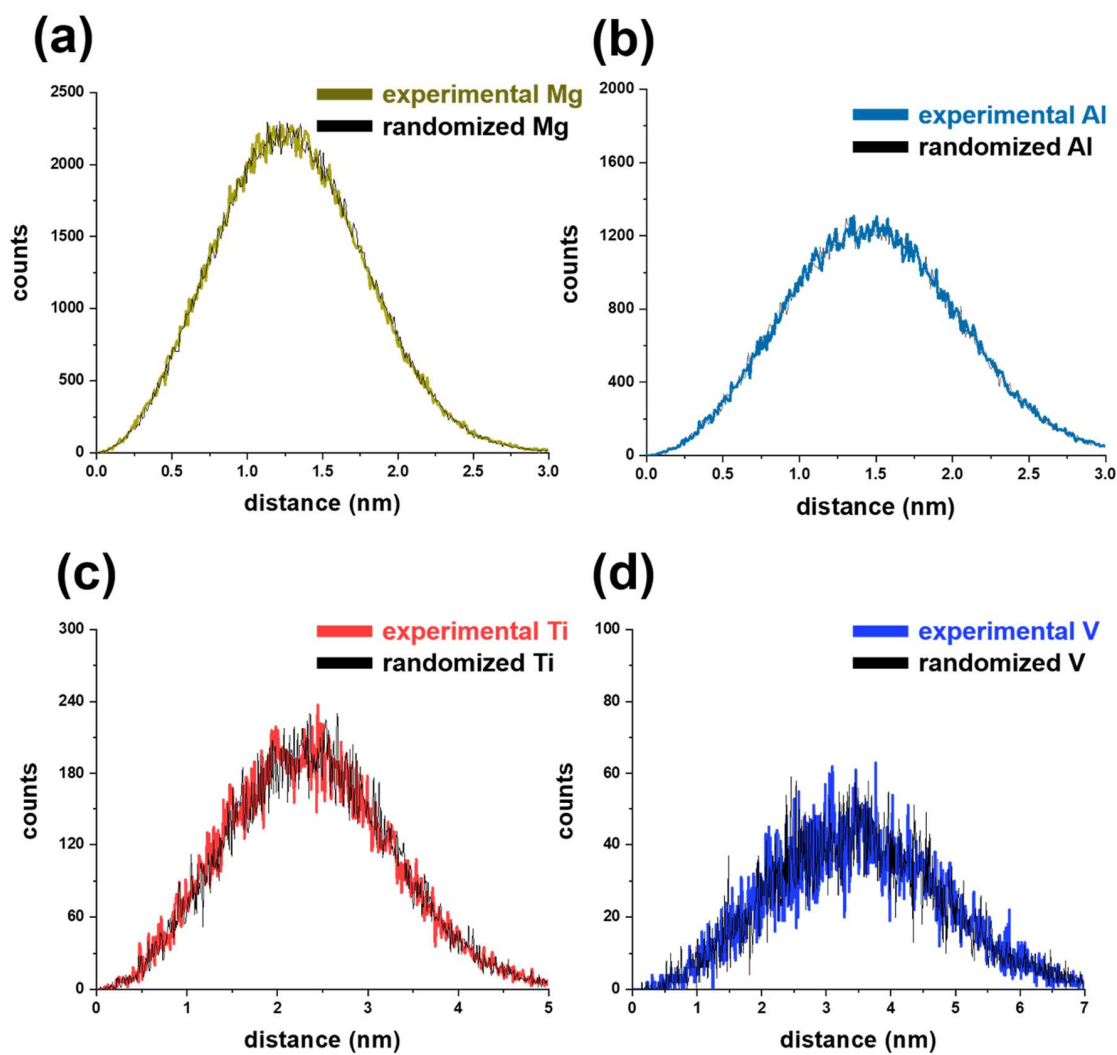

**Figure S8.** N-N nearest-neighbor atomic distance distribution of (a) Mg, (b) Al, (c) Ti, and (d) V in reconstructed hematite 3D map (Figure 6(a)) with 0.01 nm of sampling width N-pair.

In order to quantify whether any gangue-related tramp elements exhibit a tendency to cluster, we performed an N-N nearest-neighbor analysis for Mg, Al, Ti, and V from before reduction hematite ore dataset. Each distribution was compared with a randomized N element distribution, in which the atomic positions are unchanged by the mass-to-charge ratios are randomly swapped. All experimental nearest-neighbor distributions have no significant deviation from the randomized distribution.

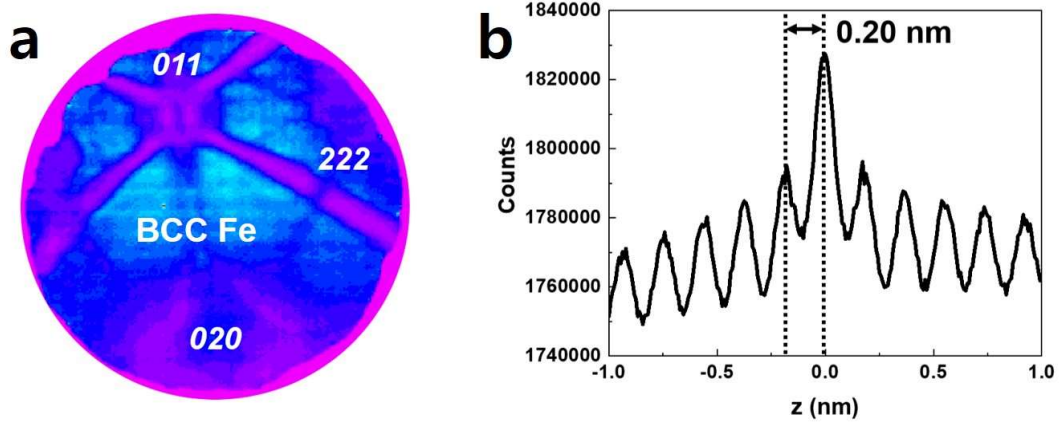

**Figure S9.** (a) Field desorption histogram from Figure 6(b). (b) Corresponding z-spatial distribution map showing the interatomic distance between Fe-Fe atoms (0.203 nm) along the <011>.

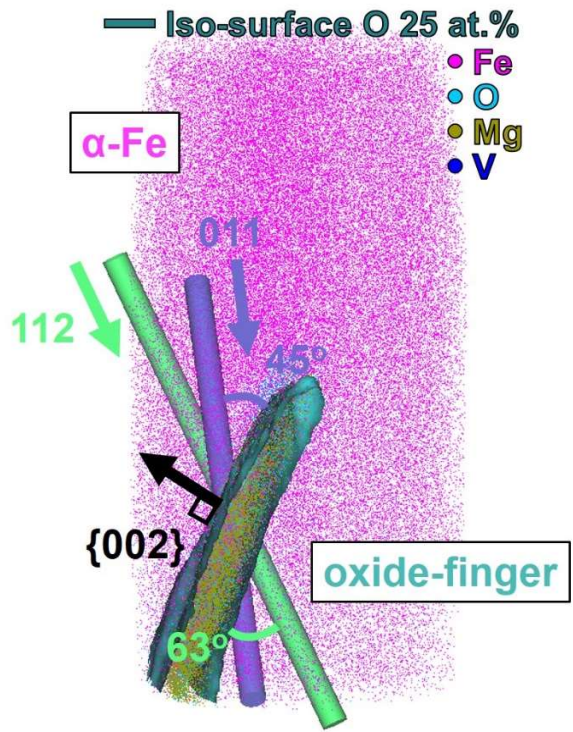

**Figure S10.** 3D atom map showing the orientation of oxides from Figure 6(c). Two poles of 011 and 112 are identified in the α-Fe region and angles between each pole and the oxide plane are calculated to be 45º and 63º which gives {002} for the oxide finger.

**Table S2.** Atomic composition of elements for before and after direct reduction of iron ore measured by APT

|  |  | Fe | O | H | Mg | Al | V | Ti | Mn | Ca | Si | Co | C | Na | Ga |
|---|---|---|---|---|---|---|---|---|---|---|---|---|---|---|---|
| **Before reduction**[I] | | 44.35 ±0.005 | 51.64 ±0.004 | 2.95 ±0.001 | 0.32 ±0.001 | 0.23 | 0.15 | 0.12 | 0.08 ±0.001 | 0.06 | 0.05 | 0.02 | - | - | 0.01 |
| **After reduction** | α-Fe[†] | 98.64 ±0.001 | 0.01 | 1.30 ±0.001 | - | - | 0.01 | - | - | 0.01 | - | - | 0.02 | - | 0.01 |
| | oxide nano-finger[‡] | 36.66 ±0.048 | 33.33 ±0.015 | 0.54 ±0.006 | 2.55 ±0.035 | 2.67 ±0.012 | 19.09 ±0.008 | 0.72 ±0.002 | 2.85 ±0.017 | 0.54 ±0.009 | 0.28 ±0.010 | 0.37 ±0.047 | 0.25 ±0.032 | 0.01 ±0.003 | 0.05 ±0.005 |
| | Mg-rich[*] | 9.04 ±0.043 | 48.85 ±0.023 | 0.30 ±0.006 | 4.04 ±0.031 | 3.46 ±0.029 | 28.20 ±0.023 | 1.22 ±0.003 | 3.55 ±0.029 | 0.14 ±0.13 | 0.28 ±0.005 | 0.5 ±0.009 | 0.25 ±0.005 | 0.01 ±0.002 | 0.15 ±0.008 |
| | Ti-rich[*] | 3.18 ±0.014 | 54.41 ±0.018 | 0.18 ±0.002 | 0.05 ±0.004 | 0.07 ±0.002 | 33.36 ±0.019 | 8.12 ±0.001 | 0.03 ±0.002 | - | 0.01 ±0.001 | 0.35 ±0.001 | 0.16 ±0.003 | - | 0.07 ±0.006 |
| | combined data[✱] | 72.19 ±0.006 | 14.83 ±0.001 | 1.00 ±0.001 | 0.46 ±0.002 | 0.52 ±0.001 | 8.49 ±0.001 | 1.48 | 0.44 ±0.001 | 0.07 ±0.001 | 0.08 | 0.14 | 0.08 ±0.002 | 0.01 | 0.21 ±0.001 |

[I]Figure 6(a). [†]Figure 6(b). [‡]Figure 6(c). [*]Figure 7(a). [✱]Merged APT data-sets of the reduced samples. Note that Ga ions are incorporated during the FIB process and that some of the H can stem from the atom probe analysis chamber and/or be incorporated during the sample transfer.

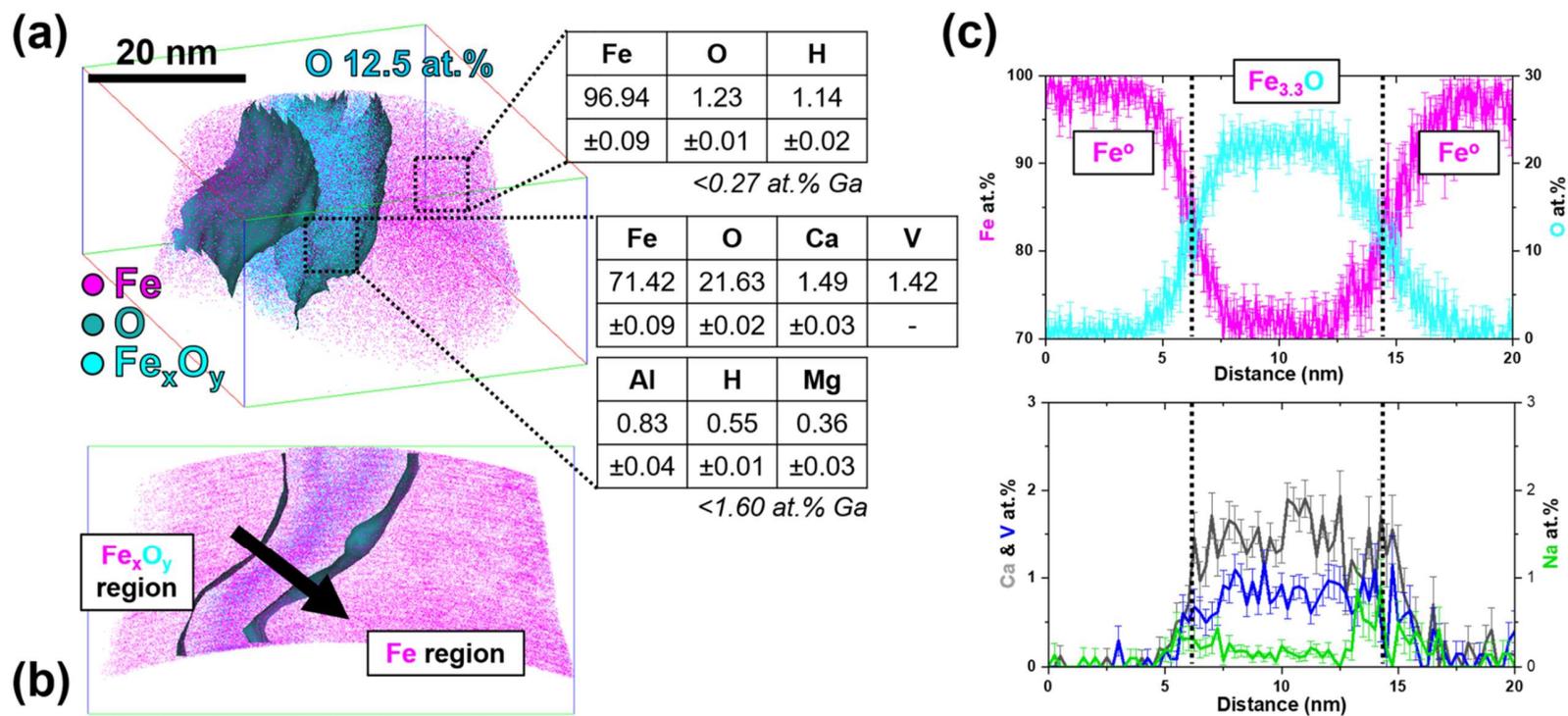

**Figure S11.** 3D atom maps of the partially reduced hematite ore (30 min, 700°C) (a) and corresponding tomogram (b) and one-dimensional composition profile from the cylindrical region of interest across the oxide nano-feature (c). All the detected gangue elements, e.g., V, Ca and Na, are enriched in the remaining oxide. The Fe to O ratio in the oxide is 3.3, which is far out of equilibrium, indicating the presence of the transient state oxide. Similar to those shown in Figure 7c, Na atoms are also accumulated at the interface between the retained oxide and the reduced matrix.